\documentstyle[preprint,aps]{revtex}
\begin{document}
\preprint{IMSc/94/30}
%

\title{Temperature dependent gap anisotropy from interlayer
tunneling}

\author{V. N. Muthukumar $^\ast$ and M. Sardar $^\dagger$ }

\address{${}^\ast$    The Institute of Mathematical Sciences,
                      Madras 600 113, India}
\address{${}^\dagger$ Institute of Physics, Sachivalaya Marg,
                      Bhubaneswar 751 005 India}

\date{\today}
\maketitle

\begin{abstract}

A recent experiment by Ma and collaborators shows that the gap
anisotropy in ${\rm Bi}_2 {\rm Sr}_2 {\rm Ca}_1 {\rm Cu}_2 {\rm
O}_{8+x}$ is strongly temperature dependent.  In particular, the
superconducting gap along the $\Gamma - M$ direction shows a weaker
temperature dependence than
the gap along the $\Gamma - X$ direction which decreases rapidly with
temperature.  We explain this novel feature as a natural consequence of
the interlayer tunneling mechanism of superconductivity.

\end{abstract}
\pacs{}
The nature of the order parameter in the cuprate superconductors is of
central importance to the study of high temperature superconductivity.
While several experiments such as IR reflectivity \cite{ir}, Raman
spectroscopy \cite{raman}, NMR \cite{nmr} and Josephson tunneling
\cite{josephson} imply that a conventional s-wave order parameter as in
BCS theory is unlikely, Angle Resolved Photoemission Spectroscopy
(ARPES) provides direct evidence that the gap in the cuprate
superconductors is highly anisotropic \cite{shen1}.  The main results
that can be inferred from low temperature ARPES \cite{shen2} are :
(i) the superconducting
gap attains its maximal value along the $\Gamma - M$ direction (Cu-O
bond direction in real space); (ii) the gap is smallest along the $\Gamma
-X(Y)$ direction (diagonal to the Cu-O bond direction) and (iii) there
is a monotonic increase in the magnitude of the gap from its smallest
value along the $\Gamma - X$ line to the largest value along $\Gamma -
M$, thereby indicating true anisotropy.

In a recent paper \cite{ma}, Ma and co-workers have presented the first
detailed analysis of photoemission spectra of superconducting Bi 2212
observed along $\Gamma - M$ and $\Gamma - X$ directions at different
temperatures.  Their results show that the gap anisotropy observed in
this material is {\em strongly dependent on temperature}, contrary to
what happens in conventional anisotropic superconductors such as Pb. In
particular they show that the gaps observed along the two high symmetry
directions obey different temperature dependences.  The gap along the
$\Gamma - M$ direction shows a very weak temperature dependence whereas
the gap along $\Gamma - X$ decreases rapidly as temperature increases.
Consequently, the gap anisotropy increases with temperature by about a
factor of 8 before falling to zero at $T_c$ .
Since these results reflect the property of the superconducting
condensate in the two different high symmetry directions,they
impose constraints on possible mechanisms of high temperature
superconductivity.

In this letter, we show that these results can be explained by the
interlayer tunneling mechanism of high temperature superconductivity
\cite{wha}.  We show that the gap equation resulting from interlayer
tunneling \cite{chak} readily distinguishes between the gaps along the
$\Gamma - M$ and $\Gamma - X$ directions and the temperature dependences
of the gaps along these two high symmetry directions are completely
different. Therefore, a temperature dependent gap anisotropy follows
very naturally from the interlayer tunneling mechanism. Our analysis
shows that the gap anisotropy increases rapidly with temperature and the
results are consistent with those reported in ref.\cite{ma}.

We begin by writing the gap equation from interlayer tunneling
\cite{chak},
\begin{equation}
\Delta_k \ = \ T_J(k) \displaystyle{{\Delta_k \over 2E_k}}\
{\rm tanh} \displaystyle{{\beta E_k \over 2}} + ~ V_{BCS}~
\sum_q{}^{\prime} \
\ \displaystyle{{\Delta_k \over 2E_q}} \ {\rm tanh}
\displaystyle{{\beta E_q \over 2}} \ .
\end{equation}
This equation can be obtained by considering two close Cu-O layers as in
Bi 2212 coupled by a Josephson tunneling term of the form,
$$
H_J~~ = ~~ -{1 \over t}
\sum_{k}~~t^{2}_{\perp}(k)~~(~c^{\dag}_{k\uparrow}
c^{\dag}_{-k\downarrow} d_{-k\downarrow} d_{k\uparrow} + {\rm h.c.}~)~~~~,
$$
where $t$ is a band structure parameter
in the dispersion of electrons along the Cu-O plane and
$t_{\perp}(k)$ is the bare single electron hopping term between
the two coupled layers $c$ and $d$.
The quantity  $T_J(k)$
in the right hand side of equation (1) is given by $T_J(k) = {t^2_{\perp}(k)
\over t}$.  The dispersion of electrons along the Cu-O plane is chosen
to be of the form
$$
\epsilon (k) ~~ = ~~ -2t~({\rm cos}k_x~+{\rm cos}k_y)~~+~~
4t^{\prime}~{\rm cos}k _x~{\rm cos}k_y~~,
$$
with $t$ = 0.25 eV and ${t^{\prime} \over t}$ = 0.45. We also choose
$\epsilon_F$ = -0.45 eV corresponding to a Fermi surface which is closed
around the $\Gamma$ - point.  These choices
are inspired by band structure calculations \cite{band}.  Note that the
Josephson coupling term in $H_J$ conserves the individual momenta of the
electrons that get paired by hopping across the coupled layers.  This is
as opposed to a BCS scattering term  which would only conserve the
center of mass momenta of the pairs.  This is the origin of all features
that are unique to the interlayer tunneling mechanism.  The second term
in the right hand side of equation (1) is obtained by postulating that
the dominant {\it in-plane} pairing mechanism is the electron-phonon
interaction.  The primed sum in equation (1) is over a shell about the
Fermi surface of width $\hbar \omega_D$, the Debye energy.

Let us first consider equation (1) in the limit of zero temperature
\cite{chak}.  We then have
\begin{equation}
\Delta_k ~~ = ~~ {T_J(k) \over 2E_k}~\Delta_k ~~ + ~~ \Delta^s~~,
\end{equation}
where $\Delta^s = V_{BCS} \sum^{\prime}_q {\Delta_q \over 2 E_q}$, is a
finite s-wave component of the gap.  Obviously,
the anisotropy in the gap at zero temperatures is
principally due to the momentum dependence of $T_J(k)$, viz., that of
$t_{\perp}(k)$.  This momentum dependence can be inferred from
electronic structure calculations.  As shown in ref.\cite{chak}, it is
adequate to choose $T_J(k) = {T_J \over 16} ({\rm cos}k_x - {\rm cos}k_y)^4$ to
reproduce the results of such calculations.  With this choice then, it is
obvious from equation (2) that the gap is smallest when $T_J(k) = 0$ (as
long as $T_J$ $>$ $V_{BCS}$).  This happens when $k_x = k_y$.  Therefore
along the $\Gamma - X(Y)$ directions where $k_x = k_y$, the gap attains
its smallest value $\Delta^s$.
The largest value of the
gap is obtained when the Fermi surface includes the points where
$T_J(k)$ is maximum, viz.,$(0,\pm \pi)$ and $(\pm \pi,0)$.  In this
case, the value of the gap is given by ${T_J \over 2} + \Delta^s$ as can
be seen from equation (2).  So, we see that the zero temperature gap from
interlayer tunneling is highly anisotropic with maxima at $(0,\pm \pi)$
and $(\pm \pi,0)$ and minima along the $k_x = k_y$ line.  These results
are in agreement with the ARPES data on gap anisotropy at low
temperatures.  We now show that this gap anisotropy increases as
temperature increases.

To see this we go back to the gap equation at finite temperature,
equation (1).  We first consider the gap at $(0, \pi)$ on the $\Gamma -
M$ line, $\Delta(\Gamma - M)$ which is given by
\begin{equation}
\Delta (\Gamma - M) ~~ = ~~ {T_J \over 2}~ {\rm tanh} {\beta \Delta
(\Gamma - M) \over 2} ~ + ~ \Delta^s(T)~~,
\end{equation}
where
$$
\Delta^s(T) ~ = ~ V_{BCS} ~ \sum_q{}^{\prime} ~ {\Delta_q \over 2 E_q}~
{\rm tanh}{\beta E_q \over 2} ~~.
$$
The temperature dependence of $\Delta(\Gamma - M)$ will be governed by
the first term in the right hand side of equation (3) since we always
choose to work in the limit of $\Delta^s(T)$ $<<$ $T_J$.  This limit
corresponds to the physical choice of $T_J$ giving rise to high
transition temperatures rather than the electron-phonon interaction
$V_{BCS}$.  Since the structure of the gap equation (3) is very
different from that of the BCS gap equation, it is clear that the
temperature dependence of $\Delta(\Gamma - M)$ will be unlike that of
a BCS gap.  For instance, $\Delta(\Gamma - M)$
will fall steeper near $T_c$ than a BCS gap would.
This is seen most simply
in the limit $\Delta^s(T) \rightarrow 0$ \cite{vnm1}.  In this limit,
the zero temperature gap $\Delta_0(\Gamma - M) = {T_J \over 2}$.  It is
readily seen that the temperature at which $\Delta(\Gamma - M)$ falls to
half its zero-temperature value is given by
${\displaystyle {T_J \over 8 {\rm tanh}^{-1}({1 \over 2})}}$.
For typical values of $t_{\perp}$ and $t$, this temperature
corresponds to $T~\sim$ 0.95$T_c$.  Consequently, we expect
$\Delta(\Gamma - M)$ to show a weak temperature dependence at low and
intermediate temperatures.  It should be emphasized that this
temperature dependence (as given by the first term in the right hand
side of equation (1)) is directly due to the ``momentum-space locality''
of the Josephson interaction $H_J$.

We have solved equation (3) for temperatures ranging from 0 to $T_c$.
Our choice of parameters are : $t$ = 0.25 eV, $t^{\prime}$ = 0.1125 eV
and $\epsilon_F$ = -0.45 ev as mentioned earlier,
$t_{\perp}$ = 0.091 eV, $V_{BCS}$ = 0.06 eV and $\hbar
\omega_D$ = 0.02 eV.  This choice of parameters leads to a purely {\it
in-plane} $T_c$ of $\sim$ 5K and a bulk $T_c$ of 83K.  The zero
temperature gap at $(0,\pi)$, $\Delta_0(\Gamma - M)$ is found to be 18.09
meV.  In fig.(1), we have shown our results for the temperature
dependence of $\Delta(\Gamma - M)$ and compared them with the
experimental results of Ma et al \cite{ma}.
Note in particular that the measured gap (a) is very
weakly temperature dependent at low and intermediate temperatures
and (b) falls very steeply near $T_c$.  As we mentioned earlier,
both these  features follow directly from equation (3) as consequences
of the Josephson interaction $H_J$.  In view of this, we suggest that
PES along the $\Gamma - M$ direction actually probes the gap resulting
from interlayer tunneling.

We now consider the gap along the $\Gamma - X(Y)$ directions where $k_x
= k_y$.  In this case, the first term in the right hand side of equation
(1) drops out and the gap $\Delta(\Gamma -X)$ is given by
\begin{equation}
\Delta(\Gamma - X)~~ = ~~ \Delta^s(T) ~~ = ~~
V_{BCS}~\sum_q{}^\prime~{\Delta_q \over
2E_q}~{\rm tanh}{\beta E_q \over 2}~~.
\end{equation}
It is obvious that the temperature dependence of $\Delta(\Gamma -~X)$
will be different from that of $\Delta(\Gamma - M)$.  Note that
equation (4) looks very much like the BCS gap equation.  However it
should be emphasized that $\Delta(\Gamma - X)$ is {\it not} a BCS gap
since the sum in equation (4) also contains the $T_c$ enhancement effects of
the Josephson interaction. In fact our results show that the gap
$\Delta(\Gamma - X)$ falls much faster from its zero temperature value
than a BCS gap does.  Therefore we get a non BCS-like temperature
dependence for $\Delta(\Gamma - X)$ as well. It is not necessary to
solve equation (4) to obtain the temperature dependence of
$\Delta(\Gamma - X)$ since $\Delta(\Gamma - X)$ = $\Delta^s$ and we have
already solved for $\Delta^s(T)$ when we obtained the temperature
dependence of $\Delta(\Gamma - M)$.  From our results
for $\Delta(\Gamma - M)$,
it is easy to see why $\Delta(\Gamma - X)$ decreases rapidly with temperature.
For instance, as $T$ increases from 0.1$T_c$ to
0.8$T_c$, $\Delta(\Gamma - M)$ decreases from 18 meV to 13 meV, we
find that $\Delta^s$ has to decrease by a factor of 3 for equation (3)
to be self consistent.  This is because of the first term in the right
hand side of equation (3) which depends weakly on temperature.
Therefore, the rapid decrease of $\Delta(\Gamma - X)$ with temperature
is actually a consequence of the weak temperature dependence of
$\Delta(\Gamma - M)$.
We have obtained $\Delta(\Gamma - X)$ for various temperatures
with the same set of parameters as before.  We find that there is a
quantitative discrepancy between our results and those of Ma et al.
With our choice of parameters, we find the zero temperature value of the
gap along $\Gamma -X(Y)$, $\Delta_0(\Gamma -X)$ $\sim$ 2 meV.  This value is
closer to the values reported in earlier ARPES experiments \cite{shen2}.
On the other hand, the low temperature value
quoted by Ma et al is $\Delta(\Gamma - X)$ = 10$\pm$2 meV at $T$ = 0.48
$T_c$.
While it is possible for us to tune $V_{BCS}$ and obtain larger values of
$\Delta(\Gamma - X)$ without altering $T_J$ and $T_c$ substantially,
we have not done so for the following reasons. The first is
that the observed values of the gaps along $\Gamma -X$ are very
sensitive to sample quality and the time elapsed between cleaving and
observation of the spectrum \cite{shen2}.  Secondly, we feel that given the
level of approximation involved in this modelling, such a fine-tuning of
parameters is unwarranted.
Instead, to see how the gap anisotropy increases with temperature, we
only consider the temperature dependence of $\Delta(\Gamma - X)$
normalized to its zero temperature value.
This is still
meaningful as we are only interested in seeing how the gap anisotropy
grows from its low temperature value and not in the absolute values of
the gaps themselves.
On comparing our results for the temperature dependence of $\Delta(\Gamma -
X)$, with those of Ma et al, we find that at low and
intermediate temperatures there is a discrepancy of 5-10\%
.  At temperatures close
to $T_c$, the discrepancy is slightly more. This is because the
experimental results indicate that the gap becomes vanishingly small at
0.84 $T_c$ whereas a mean field theory such as ours would always produce
gaps $\Delta(\Gamma - M)$ and $\Delta(\Gamma - X)$ that vanish self
consistently at identical temperatures.

In fig.(2), we have shown the temperature dependences of both $\Delta(\Gamma
- M)$ and $\Delta(\Gamma - X)$ as obtained from equation (3).
The solid line shows the temperature
dependence of the former and the dashed line shows that of the latter.
We find that $\Delta(\Gamma - X)$ falls to half its zero-temperature
value at temperatures as low as 0.6$T_c$!  The observed experimental
value of this temperature is 0.64$T_c$ \cite{ma}.  Recall that a
conventional BCS gap falls to half its zero-temperature value at $T$ $\sim
$ 0.9$T_c$.  As $T \rightarrow T_c$, we find that $\Delta(\Gamma - X)$
decreases by an order of magnitude at $T \sim$ 0.9$T_c$ while at this
temperature, $\Delta(\Gamma - M)$ has only decreased by a factor of 2.
Thus, the gap anisotropy increases from its low temperature value by a
factor of 5.

For completeness, we have also investigated the temperature dependence
of the gap anisotropy when the order parameter has mixed s- and d- wave
symmetries.  We have done this because the low temperature ARPES results
\cite{shen2} can also be explained by any model exhibiting d-wave
superconductivity \cite{pines}. However the results of ref.\cite{ma},
particularly the temperature dependent gap anisotropy cannot be obtained
from an order parameter having pure d-wave symmetry.  This is because the
anisotropy ratio of the gap for any two $k$ vectors is temperature
independent if the gap function has a purely d-wave character.  One
possibility is an order parameter with mixed s- and d- symmetries.  The
pairing potential leading to such an order parameter will be of the
form
$$
V_{kk^{\prime}} ~ = ~ V_0 + V_1 ({\rm cos}k_x - {\rm cos k_y}) ~ ({\rm
cos}k_{x^{\prime}} - {\rm cos}k_{y^{\prime}})~~.
$$
We have solved the gap equation resulting from this interaction for
various choices of $V_0$ and $V_1$ {\em without} including the
interlayer Josephson interaction.  We find that at low temperatures, the
gap anisotropy always decreases irrespective of the relative strengths
of $V_0$ and $V_1$.  At intermediate temperatures, it is possible to
obtain an increase in the gap anisotropy by fine-tuning the parameters
but this increase is only marginal.  These results underscore the
importance of the Josephson interaction $H_J$ in establishing and
enhancing the gap anisotropy.

Finally, we address the question of the sensitivity of our results {\em
vis a vis} our choice of parameters.  We have solved equation (1) for
various choices of $t_{\perp}$ and $V_{BCS}$ keeping the other
parameters fixed.  We find that as long as the value of the s-wave
component of the gap is much smaller than $T_J$, i.e., in the limit of
interlayer tunneling being stronger than any intralayer interaction, the
gap anisotropy always increases with temperature and differences in
results are only quantitative.

To conclude, we have shown that the interlayer tunneling mechanism produces
a gap anisotropy that grows with temperature.  This is because the
gap equation from interlayer tunneling leads to different temperature
dependences for the gaps along the two high
symmetry directions in Bi 2212. The gap along the $\Gamma - M$ direction
shows a much weaker temperature dependence than the gap along the
$\Gamma - X$ direction which decreases rapidly as temperature increases.
Consequently, the gap anisotropy increases with
temperature by a factor of $\sim$ 5. This is in good agreement with
experimental results that show an increase by a factor of $\sim$ 8. Our results
and those of Chakravarty et al. \cite{chak} show that the interlayer
tunneling mechanism of high temperature superconductivity can account
satisfactorily for the ARPES experiments in superconducting Bi 2212.


%
%

\acknowledgements
We thank P. Ravindran and R. Shankar for useful discussions and
suggestions.

%
%

\newpage
{\bf Figure Captions}\\
All the results were obtained with the following choice of parameters :
$t$ = 0.25 eV, $t^{\prime}$ = 0.1125 eV, $\epsilon_F$ = -0.45 eV,
$t_{\perp}$ = 0.091 eV, $V_{BCS}$ = 0.06 eV and $\hbar \omega_D$ = 0.02
eV.
\begin{enumerate}

\item
$\Delta(\Gamma - M)$ as a function of the reduced temperature ${T \over
T_c}$.  Solid line is as calculated from equation (3) in text. The experimental
data is from ref.\cite{ma}.
\item
Temperature dependence of the gaps (normalized to zero temperature
values) along $\Gamma - M$ (solid line) and $\Gamma - X$ (dashed line)
as a function of reduced temperature obtained from equation (3).

\end{enumerate}

\end{document}